%% file: main.tex
\PassOptionsToPackage{table}{xcolor}
\documentclass[sigconf,balance=false]{acmart}
\usepackage{hyperref}
\usepackage{popets}
\usepackage{enumitem}
\usepackage{float}
\usepackage{subcaption}
\usepackage{tabularx}  %
\usepackage{multirow}
\usepackage{xcolor}
\usepackage{longtable}
\usepackage{xspace}
\usepackage{pifont}
\usepackage{pdfpages}
\usepackage{url}
\usepackage[all]{nowidow} %

\newcommand{\mpixel}[1][]{\textit{Meta Pixel#1}\xspace}
\newcommand{\gpixel}[1][]{\textit{Google Tag#1}\xspace}

\newcommand{\statica}[1][]{%
  \ifx\relax#1\relax
    FDC configuration\xspace
  \else
    \if#1uFDC Configuration\xspace\else FDC configuration#1\xspace\fi
  \fi
}

\newcommand{\dynamica}[1][]{%
  \ifx\relax#1\relax
    form data collection\xspace
  \else
    \if#1uForm Data Collection\xspace\else form data collection#1\xspace\fi
  \fi
}

\newcommand{\tracker}[1][]{%
  \ifx\relax#1\relax
    tracker installation\xspace
  \else
    \if#1uTracker Installation\xspace\else tracker installation#1\xspace\fi
\fi
}

\definecolor{lightgray}{gray}{0.9}

\setcopyright{popets}
\copyrightyear{YYYY}

\acmYear{YYYY}
\acmVolume{YYYY}
\acmNumber{X}
\acmDOI{XXXXXXX.XXXXXXX}
\acmISBN{}
\acmConference{Proceedings on Privacy Enhancing Technologies}
\newlist{questions}{enumerate}{1}
\setlist[questions,1]{label=RQ\arabic*.,ref=RQ\arabic*}

\begin{document}
\title[Understanding Tracker Configuration of Form Data Collection]{Tracker Installations Are Not
Created Equal: \\Understanding Tracker Configuration of Form Data Collection}

\author{Julia B. Kieserman}

\affiliation{%
  \institution{New York University}
  \city{New York}
  \state{NY}
 \country{USA}}

\author{Athanasios Andreou}

\affiliation{%
  \institution{New York University}
  \city{New York}
  \state{NY}
 \country{USA}}

\author{Chris Geeng}

\affiliation{%
  \institution{Northeastern University}
  \city{Seattle}
  \state{WA}
 \country{USA}}
 
\author{Tobias Lauinger}

\affiliation{%
  \institution{New York University}
  \city{New York}
  \state{NY}
 \country{USA}}

\author{Damon McCoy}

\affiliation{%
  \institution{New York University}
  \city{New York}
  \state{NY}
 \country{USA}}

\renewcommand{\shortauthors}{Julia Kieserman, et al.}

\begin{abstract}
Targeted advertising is fueled by the comprehensive tracking of users' online activity. As a result, advertising companies, such as Google and Meta, encourage website administrators to not only install tracking scripts on their websites but configure them to automatically collect users' Personally Identifying Information (PII). In this study, we aim to characterize how Google and Meta's trackers can be configured to collect PII data from web forms. We first perform a qualitative analysis of how third parties present form data collection to website administrators in the documentation and user interface. We then perform a measurement study of 40,150 websites to quantify the prevalence and configuration of Google and Meta trackers.

Our results reveal that both Meta and Google encourage the use of form data collection and include inaccurate statements about hashing PII as a privacy-preserving method. Additionally, we find that Meta includes configuring form data collection as part of the basic setup flow. Our large-scale measurement study reveals that while Google trackers are more prevalent than Meta trackers (72.6\% vs. 28.2\% of websites), Meta trackers are configured to collect form data more frequently (11.6\% vs. 62.3\%). Finally, we identify sensitive finance and health websites that have installed trackers that are likely configured to collect form data PII in violation of Meta and Google policies. Our study highlights how tracker documentation and interfaces can potentially play a role in users' privacy through the configuration choices made by the website administrators who install trackers.  
\end{abstract}


\maketitle

\input{introduction}

\input{background_and_related_work}

\input{qualitative-section}

\input{quantitative-section}

\input{ethics}

\input{discussion}

\section*{Artifacts}
Artifacts used in this study, including data collection and parsing code, Google/Meta documentation, and a prototype of the Meta configuration Chrome extension are available at: \url{https://github.com/CybersecurityForDemocracy/trackers-not-equal}.
\begin{acks}
The authors used Grammarly~\cite{grammarly2025} to revise all parts of the text to correct typos, grammatical errors, and awkward phrasing. We thank Surya Mattu for discussions at the start that helped guide our study and Steven Englehardt for sharing his wisdom on large-scale tracker data collection. This work was supported by the National Science Foundation (grant numbers 2151837 and 2247516), Democracy Fund, and a gift from Reset.
\end{acks}

\bibliographystyle{ACM-Reference-Format}
\bibliography{bibliography}

\clearpage
\appendix
\input{appendix}

\end{document}

%% file: introduction.tex
\section{Introduction}
Targeted advertising is a ubiquitous marketing technique that is fueled by tracking users online to create comprehensive profiles of their activity and (presumed) interests. In order to increase the completeness of a profile, activity from an individual across devices, apps, and websites is commonly linked using ``hard identifiers'' like email addresses and phone numbers.
To collect such activity data, online advertising companies offer third-party analytics and tracking code. They encourage website administrators to include these trackers in their websites and configure them to extract Personally Identifying Information (PII) from forms that users may fill out on the website. The third party then uses this data for tracking and targeted advertising purposes.

Data extraction is of particular concern in verticals that handle sensitive consumer data, including health and finance data, and is thus subject to additional (federal) privacy regulations. In 2023, the U.S. Federal Trade Commission issued fines against two health companies, \textit{BetterHelp} and \textit{GoodRx}, for leaking sensitive data to third party tracking code providers~\cite{ftcBlogPost}. This issue was not limited to two companies --- journalists at \textit{The Markup} have uncovered instances of similar data leakage at addiction service, finance, and college preparation companies~\cite{markupAddictionArticle, markupMortgageArticle, markupChildrenArticle}.

While prior studies have analyzed leakage of PII to third parties~\cite{bujlow2017survey, chatzimpyrros2019you, englehardt2018never, senol2022leaky, dao2021alternative, starov2016you}, as well as the prevalence of trackers and data collection on sensitive websites~\cite{huo2022all, robinson2018prevalence, downing2022health, zheutlin2021data}, they have typically treated tracker installations as a binary, either installed or not installed. However, tracker installations are not created equal. They must be \emph{configured} to enable extraction of PII from forms, and to date, it is unclear how often form data extraction is enabled in practice and how the trackers' documentation and configuration user interface may influence website operators' choices.

To the best of our knowledge, no prior work explores tracker configuration across websites and the interplay between tracker documentation and real-world configurations. In this paper, we address this gap with a qualitative and quantitative approach to study how trackers can be configured for web form data collection, how these configurations are presented in the documentation and interface, and how many trackers are configured to do so in practice.

We focus on \gpixel and \mpixel, provided by Google and Meta respectively, as they are the two most popular web trackers with form data collection capabilities~\cite{ghostery}. First, we qualitatively code the documentation and configuration user interfaces of each tracker for dark patterns and other potentially confusing language (Section~\ref{sec:qualitative}). Second, we conduct a measurement study of 40,150 websites, including 3,406 health and 1,633 finance websites,
to quantify how often the two trackers are installed and configured to extract PII data from web forms (Section~\ref{sec:quantitative}).

In summary, we are guided by the following research questions. Qualitatively, we explore (i) \textit{how are form data collection features configured}, (ii) \textit{how are those configurations explained in the documentation}, and (iii) \textit{what measures do third parties take to ensure that web
administrators working with sensitive data are protecting
that data as required by US law}? Quantitatively, we investigate (iv) \textit{how prevalent is form data collection for \gpixel and \mpixel[s]}, (v) \textit{do tracker installations in health and finance verticals have different incidences of form data collection than non-sensitive verticals}, and (vi) \textit{what types of PII are trackers configured to collect}?

We find that while \textit{both Google and Meta encourage web administrators to enable form data collection by recommending the least private default configuration without addressing the potential risks}, they have significantly different interfaces. \mpixel has a streamlined set-up workflow that guides the web administrator to decisions that maximize data collection and often omits privacy considerations. \gpixel has a complex flow with contradictory statements that might make it challenging for a web administrator to assess the state of data collection configuration.

In line with this finding, our measurement reveals that \textit{\mpixel[s] are frequently configured to collect email addresses, names, and phone numbers}, either by enabling the default configuration settings that collect all supported PII fields or by custom configurations. 93.5\% of the websites that enable form data collection for Meta collect phone numbers, 93.7\% full names, and 99.5\% email addresses.
Furthermore, \textit{\mpixel[s] are more frequently configured to collect form data compared to \gpixel[s] (62.3\% vs.\ 11.6\%).}
This may be a result of Meta's aforementioned tracker configuration flow, which guides the website administrator to configure form data collection. In contrast, Google's tracker does not include form data collection as part of the setup workflow.

We also find that \textit{both Meta and Google address federal regulatory privacy restrictions on health and finance data} by requiring website administrators to indicate the vertical of their website during the account creation or tracker configuration process. However, neither provides detailed explanations that would help web administrators understand the implications of this designation.

In our measurement, we find that \textit{form data collection is less common on finance and health websites for \mpixel[s] but not for \gpixel[s]}.
Specifically, 68\% of websites with \mpixel in non-sensitive categories collect form data, compared to only 30.8\% for health and 20.3\% for finance websites, respectively.

In analyzing the configuration of trackers from multiple perspectives, we make four primary contributions:

\begin{itemize}
\item We create methodologies for qualitatively analyzing tracker configuration documentation and a data collection and analysis pipeline that enables large-scale measurement of website tracker configurations.

\item We expose how Google and Meta recommend website administrators enable automatic PII collection from form data and are providing privacy advice that has been repeatedly debunked by the US FTC.

\item We reveal that popular websites configure \mpixel to collect form data much more frequently than \gpixel.

\item We identify specific finance and health websites that have likely configured \gpixel[s] and \mpixel[s] to automatically collect PII form data in violation of Google's and Meta's policies.
\end{itemize}

We believe that our study offers a unique perspective on how instructions and setup guides can drive configuration decisions, thus providing potential technical defenses and guidance to regulators seeking to improve user privacy.

%% file: background_and_related_work.tex
\section{Background and Related Work}
\label{sec:background-and-related-work}

Online tracking is pervasive on modern websites~\cite{nikiforakis2012you, bekos2023hitchhiker}. In some cases, tracking is installed on a website through a joint effort between advertising third parties and website administrators, who maintain individual websites. These trackers take the form of a library of functions developed and made available by the third party. Not only do they record visits to websites where they are installed, but they also aim to collect Personally Identifiable Information (PII), such as email addresses or phone numbers, so that website visits can be linked to an (advertising) identity that the third party has established for a website visitor. Prior work has found that trackers do in fact facilitate the collection of website visitors' PII back to the third parties that developed them~\cite{lerner2016internet, bujlow2017survey}. This PII is collected from forms filled out by website visitors including account registration forms~\cite{chatzimpyrros2019you}, login forms~\cite{dao2021alternative}, and contact forms~\cite{starov2016you}. In some cases, PII can even be collected from a form before it has been submitted~\cite{senol2022leaky}. 

The specific type of PII collected, as well as the method of collection, varies across third-party trackers. This paper focuses on \mpixel and \gpixel as they are the two most popular web trackers with form data collection capabilities~\cite{ghostery}. \mpixel can automatically search for the following categories of user data by regular expression: email, gender, address, name, phone number, date of birth, and advertising user id (external-id)~\cite{advancedMatchingFields}. By contrast, \gpixel identifies only email addresses (also using a regular expression) but offers the option to define website-specific CSS selectors or JavaScript variables for email, address, name, and phone number~\cite{googleFirstPartyData}. Both Meta and Google trackers have an automatic and manual form data collection method; automatic collection is configured through the user interface and manual collection is configured by modifying website source code. 

\begin{figure}[ht]
    \centering
    \includegraphics[width=\linewidth]{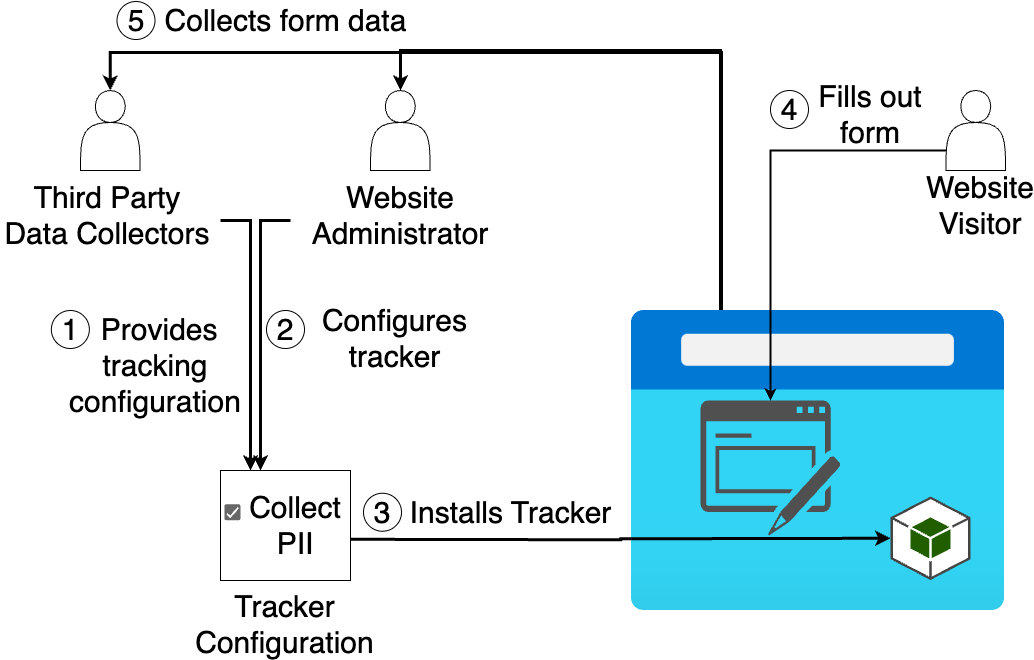}
    \caption{Tracking code setup flow.}
    \Description[Tracking code setup flow.]{A diagram that shows how data flows between third parties, website administrators, and users.}
    \label{fig:tracking-code-flow}
\end{figure}

As a concrete illustration, we discuss how \texttt{techcrunch.com}, a top-ranked news website, is configured to collect data using \mpixel. The website has a form for subscribing to an email listserv on the landing page. This flow is illustrated in Figure~\ref{fig:tracking-code-flow}.
To install the \mpixel tracking code, a website administrator had to create a Facebook account and walk through the setup steps. The website administrator may or may not be internal to the TechCrunch organization. During the process of configuring the tracking code, they would be prompted to turn on automatic data collection for several properties, including email. In this instance, \texttt{techcrunch.com}'s web admin chose to enable automatic data collection for the following PII: email, first and last name, phone number, gender, zip code, city, and state. They then installed the tracking code on their website by copying and pasting a few lines of JavaScript code from Meta's user interface.
This code will query a Meta server to load the configuration options selected by the website administrator. Now, when a website visitor decides they would like to subscribe to a newsletter, enters their email address, and clicks submit, their data will silently be passed to Meta.
Meta's tracker may also collect data in response to other click events, including other buttons on the page or anchor tags not associated with the subscription form.\footnote{This information was provided to the authors in personal communications with another researcher and verified by the authors.}

Although always privacy-invasive, this data collection becomes particularly problematic when it occurs on websites that handle legally protected data. In the United States, health and finance data are protected by the Health Insurance Accountability Act of 1996 (HIPPA)~\cite{hippa} and the Gramm-Leach-Bliley Act~\cite{grammLeachBliley}, respectively. Prior research has surfaced the role trackers have played in leaking website data in these sensitive verticals. Robinson et al. found \gpixel was installed on the majority of Illinois' hospital websites~\cite{robinson2018prevalence}, Huo et al. identified \mpixel and \gpixel data breaches across electronic health record portals~\cite{huo2022all}, and Bekos et al. found that \mpixel tracks users across time in website verticals considered sensitive by GDPR~\cite{bekos2023hitchhiker}. Additionally, reporting by news publication \textit{The Markup} anecdotally found that \mpixel installed on hospital and tax preparation websites sent website visitors' names, doctors' names (hospital websites) and income amounts (tax preparation websites) back to Meta~\cite{markupHospitalArticle, markupMortgageArticle}.

Ultimately, a tracker will only collect form data if it is configured to do so, which is determined by a website administrator at the time of installation. Throughout this paper, we use the term ``website administrator'' to refer to anyone who decides to configure and install tracking code, regardless of their technical expertise or relationship to the organization that owns the website. In practice, the decision to install a tracker and configure it for data collection may come from a combination of internal teams, like Marketing, Legal, and IT,  or be outsourced to a third-party contractor. In some cases, this may leave the original website owners unaware of the configuration setting~\cite{markupSuicideArticle}. Typically, website administrators install trackers for marketing purposes, namely to monitor website traffic, identify online customers that have followed an ad on another platform (i.e. conversions), and perform ``dynamic re-marketing''~\cite{googleRemarketing}, which targets ad campaigns to former website visitors on different platforms. 

In order to take full advantage of the marketing features of trackers, website administrators must install them such that they can collect website visitor PII. This requires that they are installed on a web page with input form fields, since trackers are scoped to a single web page, and that they are configured to use form data collection features. Configuring form data collection may not always be straight-forward to implement or understand, and as such website administrators may end up misconfiguring trackers. The consequences of misconfiguration can include legal non-compliance and data leaks. Maass et al. identified websites that had misconfigured \gpixel[s], thus violating German regulation that required anonymizing visitors' IP addresses~\cite{maass2021effective}. Prior work limited to the mobile space has further demonstrated challenges that website administrators may face in attempting to configure other kinds of privacy-impacting configurations, including working with challenging or misleading privacy APIs~\cite{zhang2024navigating} or having an incomplete understanding of third-party SDKs~\cite{alomar2022developers}.

Third parties assist web administrators with these decisions through a combination of documentation and user interface prompts in a configuration portal. Many third parties design tracking code to be primarily configured through a user interface, which allows less technically-versed people to manage the configuration. This expectation is made clear in the documentation with prompts such as ``ask your web developer'' when outlining technical steps ~\cite{googleAskDeveloper}. 

Prior research has identified the importance of quality documentation~\cite{treude2020beyond, aghajani2019software}, with a specific focus on the quality and thoroughness of privacy documentation. In the mobile space, research has shown that data collection guidance from third-party SDKs is lacking in compliance information~\cite{koch2025impact} and contains discrepancies between documented and actual behavior~\cite{inayoshi2024detection}. Even when configurations are left untouched, relying on the default configuration behavior may lead to less private behavior~\cite{rodriguez1privacy, toth2022dark}, a pitfall known as ``bad defaults'' that affects both developers~\cite{brown2021dark} and consumers~\cite{gray2018dark, mathur2019dark}.

Prior work comprehensively demonstrates the prevalence of trackers that collect form data and the challenges of documenting and configuring technical tools. However, to our knowledge, we are the first to examine and measure the specific configurations of \mpixel and \gpixel that lead to form data collection, particularly in the context of how third parties explain and guide website administrators to configure them.

%% file: qualitative-section.tex
\section{Documentation and User Interface Analysis} \label{sec:qualitative}
In order to effectively evaluate how form data configurations are explained in third-party documentation, we first needed to build an adequate understanding of how form data configurations worked. To do so, we played the role of a website administrator by creating a test website that included an input form and installing \mpixel and \gpixel as directed by the user interfaces and documentation for each tracker. Two of the paper authors then followed a reverse engineering process, modifying the tracker and observing the subsequent changes in both the form data collected and the JavaScript code loaded by the browser. We relied on this expert knowledge to evaluate how the documentation and user interface explained form data collection configurations.

\subsection{Methodology}
\subsubsection{Setup}
We reviewed 17 pages of Meta documentation and 60 pages of Google documentation produced specifically by each third party, including screenshots from the user interface. To identify documents to review, we started from the user interface, specifically focused on setup flows that did not rely on a CMS or website builder (such as Shopify or Wordpress). We then collected any documentation that was explicitly referenced or directly linked in the user interface. Finally, we included documentation we found through a Google search for how to setup the \mpixel or \gpixel or related to specific polices or best practices regarding form data collection configuration. All documentation was accessed between May and June 2024.

\subsubsection{Process}
When reviewing the selected documentation, we were guided by the following research questions:

\begin{questions}
    \item How are form data collection features configured and how are those configurations explained in the documentation?
    \item What measures do third parties take to ensure that web administrators working with sensitive data are protecting that data as required by US law?
\end{questions}

The two authors who participated in the reverse engineering process analyzed the user journey~\cite{toth2022dark} by reviewing the setup flow and documentation with a mix of deductive and inductive coding techniques. We took inspiration from existing dark patterns~\cite{mathur2021makes,toth2022dark}, software documentation~\cite{treude2020beyond}, and behavioral economics theory~\cite{kahneman2013prospect} literature. All documents were independently coded by each author before discussing. Documents were coded in an iterative process until a consensus was reached. Disagreements were resolved with the help of two additional researchers. 

\subsubsection{Codebook}
We created five codes, which are additionally enumerated in Table~\ref{table:codes} with specific examples.

\paragraph{Hidden risk}
A feature presented in such a way that a web administrator can reasonably believe they are presented with all the relevant information to make an informed choice when, in reality, additional privacy-related risks are hidden. This code takes inspiration from dark pattern ``hidden costs''~\cite{deceptiveDesignWebsite}.

\paragraph{Least private defaults}
Default configurations set by the third party that maximize data collection. Since web administrators might not modify defaults, this may lead to more data collection than would otherwise occur. This code takes inspiration from dark pattern ``bad defaults''~\cite{bosch2016tales}.

\paragraph{Least private recommendations}
Instances where third-party documentation has explicit recommendations or best practices that maximize data collection. This code was created inductively. 

\paragraph{Loss aversion}
Language that creates the perception that a web administrator would be missing out on opportunities or not getting the full offering of a certain feature by not making data maximization decisions. This code comes from the field of behavioral economics~\cite{kahneman2013prospect}. 

\paragraph{Contradictory language}
Instances where third-party guidance appears to be in contradiction to the underlying mechanisms of the data collection feature, similar to clarity issues previously explored in the context of documentation~\cite{treude2020beyond}.

\begin{table*}[t!]
    \centering
    \begin{tabular}{c p{5cm} p{5cm} c}
        \toprule
            Code & Definition & Example & Third Party \\
        \midrule
        Hidden risk & A feature presented in such a way that a web administrator can reasonably believe they are presented with all the relevant information to make an informed choice when, in reality, additional privacy-related risks are hidden & \emph{We hash the customer information on the website before they're sent to Meta technologies to help protect user privacy} & Meta \\
        \midrule
        Least private defaults & Default configurations set by the third party that maximize data collection & \emph{You have the option to enable/disable the collection of granular location-and-device data on a per-region basis. Analytics collects this data by default.}  & Google \\
        \midrule
        Least private recommendations & Instances where third-party documentation has explicit recommendations or best practices that maximize data collection & \emph{Toggle ON the parameters you want to share from your website or app. We suggest
selecting at least Email and Phone number for the best results.} & Meta \\
        \midrule
        Loss aversion & Language that creates the perception that a web administrator would be missing out on opportunities or not getting the full offering of a certain feature by not making data maximization decisions & \emph{The more items you complete on the checklist, the more complete your GA4 [Google Analytics] data will be. Many configurations determine what data is collected in your property, so it will only be available from when you complete them. That's why it's valuable to do them as soon as possible.} & Google \\
        \midrule
        Contradictory language &  Instances where third party guidance appears to be in contradiction to the underlying mechanisms of the data collection feature & \emph{Google policies mandate that no data be passed to Google that Google could use or recognize as personally identifiable information (PII)} & Google \\
        \bottomrule
    \end{tabular}
    \caption{Code book with examples observed in Google's and Meta's web tracker documentation and configuration user interfaces.}
    \label{table:codes}
\end{table*}

\subsection{Results}
\subsubsection{How are form data collection features configured, and how are those configurations explained in the documentation?}
We found that Meta and Google recommend web administrators configure \mpixel and \gpixel for data collection across the documentation and user interface and hide the possible risks of sharing web visitors' PII data with a third party. 

Meta and Google often recommend the least private configuration option, which might create hidden risks for the website administrator. For example, across the documentation, Meta recommends that web administrators use both manual and automatic data collection (as defined in Section ~\ref{sec:background-and-related-work}) to increase the amount of data collected for ``maximum performance''~\cite{facebookMaxPerformance}. Similarly, Google recommends that website administrators install a \gpixel ``on every page of your website'' rather than consider what pages are most appropriate for its intended purpose~\cite{googleMultiplePages}. Since these recommendations come directly from the third parties, it is reasonable for a web administrator to take them as a form of best practice, even though they present a least private approach.

Google and Meta also insufficiently address the risks of configuring form data collection. Both Google and Meta's documentation contains language implying that hashing is sufficient for privacy, a claim that has been debunked both by researchers and by the US Federal Trade Commission (FTC) twice in the past twelve years~\cite{demir2017pitfalls, ftcHashOld, ftcHash}. Meta's documentation states that they ``hash the customer information on the website...to help protect user privacy''~\cite{facebookHash}. However, by using Meta's preferred hashing function (SHA-256)~\cite{facebookSHA}, Meta can link customer data collected from the website to an existing Meta profile. Therefore, by omitting any other details, this language hides the risk of configuring \mpixel to share visitors' PII with Meta. Similarly, Google refers to ``sending hashed first party conversion data from your website to Google'' as sending data ``in a privacy safe way''~\cite{googleHash}.

We found that the documentation for both providers contains the same language multiple times across different setup guides. For example, Meta states their preferred matching setup, ensuring that a page has ``form fields'' that collect PII information, three different times~\cite{facebookHash, metaAutomaticEvents, facebookMaxPerformance}. Google states a preference for receiving user email addresses to help identify customer leads five times across two documents~\cite{googleEnhancedConversions, googleEnhancedForWeb}. This duplication could reinforce the effect on the reader and, since the documentation may not be accessed or read in a specific order, increase the opportunity for someone installing the tracking code to encounter these recommendations.

Although both providers generally encourage configuring data collection, we did find some discrepancies between how Meta and Google describe and present data collection configuration.

\paragraph{\mpixel}
Meta provides website administrators with a setup workflow that heavily recommends configuring data collection. When installing a new \mpixel through the basic setup flow, a wizard includes a screen dedicated to automatic data collection, described as using ``information that your customers have already provided to your business'' (Figure~\ref{fig:meta-UI}). Including a prompt to turn data collection on in the setup flow can be read as a least private recommendation by Meta to turn it on, insinuating that data collection is part of normal \mpixel use. They further omit any descriptive text about privacy considerations related to PII data collection, creating a hidden cost for web administrators who may not consider the risks without further information. Further, if the web administrator takes the prompt, all of the attributes Meta can collect are automatically selected. By assuming the web administrator would like to collect all possible attributes, Meta has set up a least private default that relies on the web administrator to actively deselect attributes to increase customer privacy (Figure~\ref{fig:meta-auto-select}). 

The documentation also mentions inconsistent behavior that may occur if a website has multiple \mpixel[s] installed~\cite{facebookMultiplePixels}. If one pixel collects form data and another pixel does not, both pixels may collect form data collection when one is triggered to do so. Although it is explicitly mentioned in the documentation, it is not included as part of the setup flow nor does the documentation adequately explain the implications for potential data leakage. 

Data submission events also behave in unexpected ways.
When automatic form data collection is enabled, Meta's data collection logic activates on any button click event, whether or not the button is connected to a form with data~\cite{metaButtonClick}. This could create instances of form leakage behavior, not just of an email and password, as documented by Senol et al.~\cite{senol2022leaky}, but also of other supported PII types such as name, address, or gender. We further found instances where a form data collection event was triggered by a completely different type of click event, including clicking on a hyperlink, which was not specified in the documentation. 

\paragraph{\gpixel}
Google's tracking ecosystem is complex. Form data collection is configured across several different parts of the user interface and \gpixel can send data to different Google products, including Google Ads and Google Analytics. While this complexity does not necessarily increase the likelihood that a website administrator will configure form data collection, it can make it hard for website administrators to determine if a \gpixel is configured to collect form data. Neither the documentation nor the user interface make it clear how to ensure that tracking code will behave as intended. Unlike \mpixel[,] the data collection feature is not explicitly mentioned in the tracker set-up flow and thus requires self-direction to turn on.

We also found that the documentation uses contradictory statements. For example, Google asserts that ``Google policies mandate that no data be passed to Google that Google could use or recognize as personally identifiable information (PII)''~\cite{googleBestPracticesPII}. However, other parts of the documentation make it explicit that tracking code ``uses first-party user-provided data from your website''~\cite{googleEnhancedConversions}. Although this discrepancy is likely drawing a distinction between hashed and un-hashed PII, this distinction is not made explicit (and, as stated, hashing is insufficient as a privacy method), therefore this can be read as a direct contradiction in the documentation.

\subsubsection{What measures do third parties take to ensure that web administrators working with sensitive data are protecting that data as required by US law?}
In the United States, sensitive data in health and finance verticals are provided with specific legal protections. Both Meta and Google address regulatory restrictions by applying constraints on automatic form data collection for websites belonging to either of these verticals. However, a website's vertical classification is entirely self-reported and neither Google nor Meta offer reasonable explanations that would help a web administrator understand the importance of the designation. 

Meta states that ``businesses...may not have certain features available to them if they’re categorized as being in a restricted vertical''~\cite{metaRestrictedVerticals}, omitting that the reason for this is because they presumably deal with especially sensitive data. They further state that businesses ``learn how to set up [data collection] manually''~\cite{facebookMaxPerformance}, suggesting these restrictions can be circumvented by using manual data collection techniques. This may result in a web administrator modifying the website source code to send the same data to the third party that was intentionally restricted in the user interface. Similarly, \gpixel states that data sharing ``is not available to Analytics accounts with properties in the “Health” and “Finance” property industry categories''~\cite{googleVertical} without further explanation.

\subsubsection{Takeaway}
In summary, we observed the following:
\begin{itemize}
    \item Both Meta and Google recommend configuring data collection without sufficiently addressing the privacy risks of doing so.
    \item Meta's basic setup flow prompts website administrators to configure data collection and, when configured, collects 11 types of PII by default.
    \item Both Meta and Google technically restrict websites in sensitive verticals, but vertical identification is at the discretion of the website and not adequately explained in the documentation.
\end{itemize}

%% file: quantitative-section.tex
\section{Website Configurations}
\label{sec:quantitative}
Thus far, we have analyzed how the \mpixel and \gpixel present form data collection to website administrators in the documentation and user interface. We proceed by measuring website configurations to understand what impact this may have had on actual websites.

\subsection{Methodology}
\label{sec:methodology}
In order to compare the configurations of websites in health, finance, and other verticals, we needed to generate a list of top websites by vertical. To do so, we joined the top one million websites from Tranco~\cite{LePochat2019} with SimilarWeb, a web traffic estimator that categorizes websites by vertical (e.g., finance, games, health, shopping, travel)~\cite{similarweb} and has been used by prior studies~\cite{xavier2024web, suksida2017study, similarwebUkraine, zheutlin2021data}. This generated a list of 42,481 websites with quartile ranks of 34,429, 110,756, 355,814, and 999,979 for quartiles 1, 2, 3, and 4, respectively.

\begin{figure}[!htb]
        \centering
        \includegraphics[width=\linewidth]{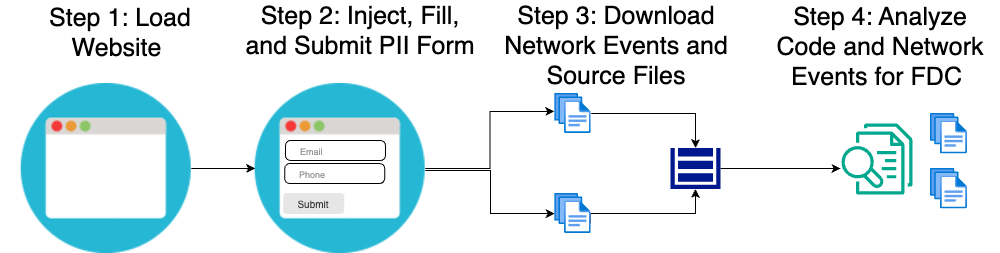}
        \caption{Data collection and analytics pipeline.}
        \Description[Data collection and analytics pipeline.]{A diagram that depicts how data was scraped and parsed.}
        \label{fig:data-crawl}
\end{figure}

To measure data collection configurations on each website, we simulated form data collection, scraped website tracker code and collected network traffic, and then parsed the data to detect the presence of trackers (i.e. \textit{tracker installation}) and whether they were configured to collect form data (i.e. \textit{form data collection (FDC)}). The pipeline is outlined in Figure~\ref{fig:data-crawl}.

\subsubsection{Data Collection}

On each website we performed the following data collection actions:
\begin{enumerate}
    \item \textit{Form Injection Action:} Inject a form with PII fields into the page and submit it. If a tracker is configured to collect form data, it will generate a network event that sends these data to Google or Meta.
    \item \textit{Download Action:} Capture all the JavaScript files and network events loaded by the webpage. This collects any potential \gpixel and \mpixel configuration files present and any form data collection events that might have occurred as a result of the form injection action.
\end{enumerate}

We note that we visited only the website landing page, which may have a smaller incidence of tracker installations than other pages on certain types of websites~\cite{huo2022all}. We chose not to interact with cookie consent banners present on the page to avoid bias towards any one user behavior (accept or decline). We address the potential impact of this decision in Section~\ref{sec: dataset-overview}.

\paragraph{Instrumentation} All website visits took place on Linux virtual machines in Google Cloud's US Central region between September and November 2024. We ran a total of 50 Virtual Machines (VMs) in parallel with identical configurations and a different subset of our website list. 

Each virtual machine had a Python script that would start on machine boot and iterate through a subset of our input website list. We used Google Cloud infrastructure (buckets and functions) to maintain the visit status of each website and communicate with individual VMs. The Python script opened each website in Chrome for 180 seconds to provide sufficient time for the page to load and our website visit actions to complete. In order to reduce the possibility of bot detection, we chose not to rely on commonly used automation tools like Selenium or OpenWPM~\cite{englehardt2016online} and restarted each machine every three hours.

Each website was opened in a Chrome instance with two installed custom Chrome extensions. The extensions performed two actions: a form injection action and a download action. The tasks were triggered after the page finished loading, determined using Chrome's manifest v2 webNavigation feature~\cite{chromeWebNav}. 

\paragraph{Form Injection Action} On each website, we injected an HTML form into the webpage with several PII fields; namely email, phone number, first and last name, city, state, and zip code. Our extension filled out the form fields with placeholder data and then triggered the submit button. The form was attached to the first \verb|<div>| or \verb|<span>| node found on the page. Our extension only injected into the top document of the page and not into nested iframes. The intent of this action was to trigger a form data collection event if a tracker on the page was configured to do so.

We opted to use our own form instead of one present on the page, as we were interested in measuring the tracker configuration, which is independent of the current HTML elements on the page. This method enabled us to measure the configurations of all websites, regardless of whether they currently have a PII form on the landing page. Consequently, we are not measuring acutal observed PII leaks but rather whether a website is configured in a way that could potentially cause PII leaks.

\paragraph{Download Action} On each website, we downloaded all the page source files, including the JavaScript and HTML files. We also downloaded all network traffic detected during our website visit, both through the network HAR file and using Chrome's manifest v2 webNavigation feature~\cite{chromeWebNav}. This ensured we captured any form data collection events potentially triggered by our form injection action and the JavaScript configuration files of any \mpixel and \gpixel tracker files loaded by the website.

\paragraph{Dataset Collection Retries}
After our initial scraping, we identified the subset of websites that either had no instances of \gpixel or \mpixel \tracker[s], or loaded a \gpixel or \mpixel but had no detected instances of \dynamica, using a methodology described in the subsequent subsection. To reduce possible false negative detections, we then performed another round of website visits, including form injection, within this subset of websites. We repeated this process until every website had been visited at least three times or had a detected instance of \dynamica. The diminishing returns are enumerated in Section \ref{sec: dataset-overview}.

We analyzed the data across all visits with the following parsing logic.

\subsubsection{Data Parsing}
After each scrape, we processed all collected files (i.e., network traffic and downloaded source code) through our parsing pipeline. Our parsing logic allowed us to answer three questions: (1) Did this website have a \mpixel and/or \gpixel installed? (2) Did the \mpixel and/or \gpixel perform \dynamica with our placeholder form data? (3) Was the \mpixel configuration file configured to collect form data?

\paragraph{Detecting \tracker[u]}
We used network GET requests to determine if a website had a \mpixel or \gpixel installed. The specific URLs used to load each tracker were identified during the reverse engineering process. We determined if a website had a \mpixel installed by parsing \mpixel configuration files, loaded by a GET request to \textit{\url{connect.facebook.net/signals/config/}}. An example is shown in Figure~\ref{fig:meta-get-request} in the appendix.

Similarly, we determined if a website had \gpixel installed by parsing \gpixel configuration files, loaded by a GET request to \textit{\url{googletagmanager.com}} (see Figure~\ref{fig:google-get-request} in the appendix). There are four types of tags that feed data to different parts of the Google ecosystem but are all considered valid \gpixel installations. They can be identified by their tag ID prefix: \texttt{AW} (Google Ads), \texttt{DC} (Google Floodlight), \texttt{G}/\texttt{GT} (Google Analytics 4), and \texttt{UA} (Universal Analytics).

In addition, Google supports first-party mode~\cite{googleFirstPartyMode}, which allows website administrators to install infrastructure that sits between a website and Google's servers. In these cases, \gpixel will be loaded from a domain chosen by the website administrator instead of Google, but still retain the \textit{\url{googletagmanager}} file name. We consider these valid installations if they match the structure of a standard \gpixel configuration file.

\paragraph{Detecting Form Data} \label{sec: analysis-types}
To measure instances of form data collection, we performed both a dynamic and a static analysis. The dynamic analysis allowed us to determine if \gpixel[s] and \mpixel[s] collected form data, and the static analysis allowed us to determine if and how \mpixel[s] were configured for form data collection (unfortunately, we were unable to directly determine how \gpixel[s] were configured).

In our \textbf{dynamic analysis}, we leveraged the output produced by our form injection action; we parsed captured network traffic %
and detected form data collection through known tracker URLs with query parameter values containing the (hashed) placeholder data we submitted through the form injection task. These URLs were identified in the initial reverse engineering process, and are listed in Table~\ref{tab:collection-urls}. We identified our data by hashing it according to the standard for each tracker (SHA-256 for \mpixel and SHA-256 on a base64 encoded string for \gpixel) and then searching network events for a matching string hash.

If we found any instance of our hashed PII data in a request to the known tracker URLs, we labeled the \gpixel or \mpixel tracker as a website with form data collection (FDC). An example of a \mpixel \dynamica network event and \gpixel \dynamica network event are presented in Appendix Figures~\ref{fig:meta-fdc-event} and ~\ref{fig:google-fdc-event} respectively. The highlighted portion is a hashed version of the placeholder email address submitted through our form and detected by our parser.

In our \textbf{static analysis}, we parsed the \mpixel tracker files collected during the download task to identify if the tracker was configured for data collection. Through our reverse engineering process, we were able to identify where the user interface configuration choices, described in Section \ref{sec:qualitative}, appeared in the JavaScript files loaded by the browser. This analysis can determine not only whether a \mpixel is configured to collect data but also which of the 11 supported PII fields it was configured to collect, all of which are selected by default. Screenshot Figure~\ref{fig:meta-config} in the appendix illustrates how this configuration appears in the JavaScript code, identified as \textit{selectedMatchKeys}.

Unfortunately, the Google ecosystem is more complex, involves more components, and has been intentionally obfuscated, making static analysis more challenging. In addition, \gpixel form data collection is limited to email; all other PII field collection requires further manual customization from website administrators. Therefore, we leave static analysis of \gpixel code for future work.

\begin{table}[ht]
\centering
\begin{tabular}{l l } 
\toprule
\textbf{Third Party} & \textbf{Data Collection URLs} \\ \midrule
Meta & facebook.com/privacy\_sandbox/register/trigger \\ 
      & facebook.com/tr\\ \midrule
Google & googleadservices.com/pagead/conversion  \\ 
      & google.com/ccm/form-data/ \\
      & analytics.google.com/g/collect \\
      & google.com/pagead/form-data/ \\ \bottomrule
\end{tabular}
\caption{Tracker URLs that receive form data from \mpixel and \gpixel installations on websites.}
\label{tab:collection-urls}
\end{table}

\paragraph{\dynamica[u]: Automatic vs. Manual}
As mentioned in Section \ref{sec:background-and-related-work}, \mpixel and \gpixel support two modes: manual (when a website administrator writes JavaScript to send PII fields to the third party explicitly) and automatic (when a website administrator authorizes the third party to auto-detect PII fields through the user interface).

We were able to differentiate between \mpixel automatic and manual \dynamica by the keys specified in the network event, determined through trial and error during the reverse engineering process. Automatic form data collection uses the key \textit{udff}, followed by an abbreviation of the PII type. For example, email configured through the user interface will appear in a url as \textit{\text{{udff[em]}}}, followed by the hashed email, as demonstrated by the highlighted text in Appendix Figure \ref{fig:meta-fdc-event}. Manual form data collection uses the key \textit{ud} but otherwise appears the same way, i.e. with the PII abbreviation (or unabbreviated \textit{external\_id}) in brackets immediately following the identifier. The majority of websites with detected trackers (90.92\%) exclusively used automatic mode. The remaining websites used a combination of the two and, when manual mode was used, it was primarily for the external ID field (92.82\%).\footnote{The External ID represents users in an advertising system~\cite{metaExternalId}.}

We were not able to make the same distinction for \gpixel, as the automatic and manual network events appeared identical. However, this distinction is less meaningful for \gpixel than it is for \mpixel, as both manual and automatic modes require some interaction with the \gpixel configuration UI, and in some cases, \gpixel will default to automatic mode if manual data is not detected~\cite{googleUserProvidedData}. Therefore, we do not distinguish between the automatic and manual modes of \gpixel in our analysis.

In the rest of the paper, we refer to \tracker[s] for which we detected a PII collection event as \textbf{\dynamica[u],} and to \mpixel[s] for which our static analysis identified JavaScript data collection configuration as \textbf{\statica[u]}. We use our \dynamica analysis to report comparative measures on the differences between \mpixel and \gpixel. We use our \statica analysis to explore the specific PII data fields that \mpixel[s] are configured to collect.

\input{dataset}

\subsection{Analysis} \label{sec: analysis}
\begin{table*}[t]
  \centering
  \small
  \begin{tabular}{l r rrr rrr}
    \toprule
    \multirow{3}{*}{\normalsize\textbf{Vertical}} &  \multirow{3}{*}{\normalsize\textbf{Websites}} & \multicolumn{3}{c}{\normalsize{\textbf{Google}}} & \multicolumn{3}{c}{\normalsize{\textbf{Meta}}}  \\  
      & & \multicolumn{2}{c}{\textbf{\tracker[u]}}  &  \multicolumn{1}{c}{\textbf{\dynamica[u]}}  & \multicolumn{2}{c}{\textbf{\tracker[u]}}   &  \multicolumn{1}{c}{\textbf{\dynamica[u]}} \\  
      & & Websites & Vertical Websites  & \tracker[u]  & Websites &  Vertical Websites  &  \tracker[u] \\
    \cmidrule(lr){1-2}  \cmidrule(lr){3-5}  \cmidrule(lr){6-8} 
    \rowcolor{lightgray} \textbf{Non-Sensitive} & 35,113 & 25,471 & 72.5\% & 11.5\% &9,731  &27.7\%  & 68.0\% \\
    \textbf{Health}& 3,406  & 2,565 & 75.3\% & 11.6\%  & 1,075  & 31.6\%& 30.8\% \\
    \rowcolor{lightgray} \textbf{Finance} & 1,633  & 1,103  &67.5\%  & 13.4\%  & 503  & 30.8\% & 20.3\% \\
    \textbf{Total} & 40,150 &  29,137 & 72.6\% & 11.6\% & 11,309 & 28.2\% & 62.3\% \\
    \bottomrule
  \end{tabular}
  \caption{Breakdown of \tracker[s] and \dynamica for Google and Meta on different verticals.}
  \label{tab:vertical_breakdown}
\end{table*}

We visited popular websites to measure the configuration and form data collection of \tracker[s]. We guide our analysis of these measurements with the following research questions:

\begin{questions}[resume]
    \item How prevalent is \dynamica for \gpixel[s] and \mpixel[s]?
    \item Do \tracker[s] in health and finance verticals have different incidences of \dynamica than non-sensitive verticals?
    \item What types of PII are \tracker[s] configured to collect?
\end{questions}

\subsubsection{How prevalent is form data collection for \gpixel[s] and \mpixel[s]?}

Table~\ref{tab:dataset_overview} shows a breakdown of \tracker[s] and \dynamica[s] for Meta and Google. 

We detected a \gpixel installation on 72.6\% of websites, and a \mpixel on 28.2\% of websites. There was considerable overlap between the two. When there was a \mpixel, 98\% of the time there also was a \gpixel; only 0.6\% of the websites we studied have only a \mpixel.
In other words, \gpixel was by far the most present tracker, with a large drop-off to \mpixel.

However, \dynamica was much more prevalent on \mpixel than \gpixel. We detected a \mpixel \dynamica event on 62.3\% of websites with a \mpixel (17.6\% of all websites). In contrast, only 11.6\% of websites with \gpixel exhibited \dynamica (8.4\% of all websites).
That is, when a \mpixel is present, it is much more likely to perform \dynamica than in the case of \gpixel.
This observation is consistent with our analysis in Section~\ref{sec:qualitative}, which shows that Meta recommends the use of automatic data collection and provides a configuration UI flow that requires the website administrator to actively accept or decline the configuration of this feature. By contrast, while Google also recommends the use of automatic data collection, the required setup flow does not include a prompt to configure data collection, which is off by default.

We also observe that when there is a \mpixel on a website, \gpixel is more likely to collect form data. Table~\ref{tab:breakdown_subsets} shows that out of the 18,054 websites that have \gpixel but no \mpixel, only 5.4\% (976) perform Google \dynamica, while out of the 11,083 websites that have both trackers installed, 21.7\% (2,401) have \gpixel \dynamica. This absolute as well as relative increase indicates a correlation between websites that install \mpixel and their configuration of \gpixel to collect form data. 

We also find this correlation in the additional Logistic Regression analysis we performed, found in Appendix Table~\ref{tab:logistic_regression}, which suggests much higher odds of having Google \dynamica when there is a \mpixel (4.839), as well as moderately higher odds of having Meta \dynamica when there is a \gpixel (1.903).

\input{table_subsets}

\subsubsection{Do tracker installations in health and finance verticals have different incidences of form data collection than non-sensitive verticals?}

Table~\ref{tab:vertical_breakdown} breaks down the Google and Meta \tracker[s] and \dynamica across health, finance, and all other non-sensitive verticals. Although other categories may be considered sensitive, such as adult or gambling websites, we differentiate between health and finance specifically as they are subject to special legal restrictions in the United States (HIPPA and Gramm-Leach-Bliley Act, respectively) and are restricted from enabling automated form data collection by Meta and Google. 

We notice that  \gpixel and \mpixel \tracker[s] are consistent across verticals, with \gpixel found on 67.5\% to 75.3\% of each vertical's websites, and \mpixel between 27.7\% and 31.6\%, respectively.
\gpixel \dynamica is consistent across verticals, ranging from 11.5\% of \tracker[s] to 13.4\%. However, there is a big difference in \dynamica for \mpixel.
While 68\% of \mpixel[s] in non-sensitive verticals collected form data, only 30.8\% of health and 20.3\% of finance websites with \mpixel installations did. This suggests that Meta is somewhat effective at preventing \dynamica in these sensitive verticals. These results are further corroborated by our Logistic Regression Analysis presented in Appendix~\ref{sec:appendix-regression}, which shows an inverse relation between health/finance verticals for \mpixel \dynamica[,] while it shows no statistical significance between the same verticals and \gpixel \dynamica[.]

As discussed in Section~\ref{sec:qualitative}, both Meta and Google prohibit \dynamica on websites that identify as a health or finance website. However, that identification is at the discretion of the website administrator and, as discussed in Section \ref{sec:qualitative}, not clearly explained by either party. When reviewing our data, we found several websites configured for \dynamica that clearly belonged to health or finance. Therefore, for reasons that we did not investigate, these websites likely have an incorrect vertical classification that circumvented the third party's technical restrictions. We provide a few examples of such websites by vertical and third party.

\paragraph{Health \mpixel} (i) \textit{Avenues Recovery}~\cite{avenuesRecovery}, a network of drug and alcohol rehabilitation centers, (ii) \textit{Benefits Checkup}~\cite{benefitsCheckup}, an information service that connects seniors to various food, housing, and medical assistance programs, and (iii) \textit{Nugg MD}~\cite{nuggMd}, a medial marijuana card provider.

\paragraph{Health \gpixel} (i) \textit{Cross River Therapy}, a therapy service for children with Autism~\cite{crossRiverTherapy}, (ii) \textit{Banner Health}, a non-profit healthcare system~\cite{bannerHealth}, and (iii) \textit{National Alliance on Mental Illness (NAMI)}, a mental health organization with local support services~\cite{nami}. After our disclosure, NAMI made the necessary modifications and is no longer configured
for form data collection.

\paragraph{Finance \mpixel} (i) \textit{After Pay}~\cite{afterPay}, a buy now pay later loan company, (ii) \textit{Patriot Software}~\cite{patriotSoftware}, a small business payroll and accounting software company, and (iii) \textit{KB Card}~\cite{kbCard}, a credit card service company.

\paragraph{Finance \gpixel} (i) \textit{Equifax}~\cite{equifax}, a credit reporting agency, (ii) \textit{Capital One}~\cite{capitalOne}, a bank holding company, and (iii) \textit{Nationwide}~\cite{nationWide}, an insurance and financial services company.

Although Meta and Google's technical restrictions are a step in the right direction, they are clearly insufficient at preventing all websites in health and finance verticals from using automatic data collection features and ensuring compliance with Google's and Meta's policies.

\subsubsection{What kind of PII are \tracker[s] configured to collect?}
When analyzing \mpixel configurations, we were able to identify the specific types of PII that \mpixel[s] are configured to collect. We omit \gpixel because it can only be configured to collect email addresses automatically and requires specifying CSS or JavaScript selectors for other PII fields, which is a more technically involved task. 

\mpixel supports the following eleven PII types (some of which are grouped together): email, phone number, first and last name, city, state and ZIP code, gender, country, date of birth and external (advertising) ID. Figure~\ref{fig:meta-auto-select} in Appendix~\ref{sec:appendixa} demonstrates Meta's PII field collection interface in the setup flow.

In Section~\ref{sec:qualitative}, we found that Meta's default pre-selection of all 11 PII fields to be a case of the \textit{Least Private Defaults} pattern.
Here, we analyze how often website administrators change these default settings based on the \statica[s] we have collected and reverse-engineered for \mpixel. We limit our results to websites with only one \mpixel installed (86.5\% of all websites with \mpixel), as multiple \mpixel[s] with different configurations can have unexpected behavior and cause inaccurate tracking, as reported by Meta~\cite{metaMultiplePixel}. 

We found over half of these websites (51.3\%) used the default configuration, which enables a \mpixel to detect all PII fields. Table \ref{fig:match-key-table} shows the percentage of websites that modified the default configuration to collect specific fields.\footnote{Our analysis of PII field selection by websites in the health and finance verticals roughly follows a similar pattern to the entire dataset.} When website administrators customize their configuration, they most frequently exclude the external id, date of birth, and country fields (found in only 5.4\%, 5.0\%, and 4.6\% of custom configurations, respectively). In turn, email address, first and last name, and phone number were the most likely to be collected (48.2\%, 42.4\%, and 42.2\%). This is aligned with Cui et al.'s finding that email was the most commonly collected PII field from website forms across categories~\cite{cui2024understanding}.

Considering both default and custom configurations, the overwhelming majority of \mpixel[s] with \statica are configured to collect email, name, and phone number (99.5\%, 93.7\%, and 93.5\%, respectively). We note that these three fields, especially combined, can likely identify a specific individual. Recall that Meta documentation recommends website administrators at least toggle ON collection for email and phone numbers (Table~\ref{table:codes}). Again, we observe consistency between real-world configurations of \tracker[s] and the instructions in the documentation. 
 
\begin{table}
    \centering
    \small
    \begin{tabular}{cl c}
        \toprule
        \multicolumn{2}{l}{\textbf{Field}} & \textbf{\% \statica[u] Websites} \\
        \midrule
        \multicolumn{2}{l}{\textbf{Default  (All Fields)}}& 51.3\% \\
        \midrule
        \multirow{8}{*}{\rotatebox{90}{\large{\textbf{Custom}}}} & Email & 48.2 \% \\
        & First and Last Name & 42.4\% \\
        & Phone Number & 42.2\% \\
        & City, State, and ZIP Code & 38.5\% \\
        & Gender & 37.0\% \\
        & External ID & 5.4\% \\
        & Date of Birth & 5.0\% \\
        & Country & 4.6\% \\
        \bottomrule
    \end{tabular}
    \caption{Breakdown of PII field collection configurations of \mpixel[s.] Percentages split by default and custom configurations, and relative to all websites with exactly one \mpixel configured for form data collection. 51.3\% of these websites use the default configuration, which collects all fields.}
    \label{fig:match-key-table}
\end{table}

\subsection{Limitations}
\noindent\textbf{Geography.} All data scraping in this study was done from Google Cloud Linux machines in the United States. Therefore, any legal protections provided in other countries, such as the European Union's GDPR, are not taken into account in this study.

\noindent\textbf{Trackers.} This study focused on two trackers, \mpixel and \gpixel. Future work could investigate similar configurations on other popular trackers.

\noindent\textbf{Privacy Policies.} This study did not look at privacy policies and thus did not measure the disclosure of PII form data collection to website visitors in those policies.

\noindent\textbf{Landing Page Visits.} This study only visited the landing pages of websites. For some websites~\cite{huo2022all}, \tracker[s] might be less frequent on their landing pages compared to their other pages.

\noindent\textbf{Webpage Forms.} As demonstrated in this study, PII form data collection requires a form with at least one PII field. Our methodology used a generated form to measure whether data collection was \emph{enabled}. We did not measure if a website actually had a PII form or if a website had real PII data leaks.

%% file: dataset.tex
\subsection{Dataset Overview and Validation}
\label{sec: dataset-overview}

\begin{table*}[t]
    \centering
    \small
    \begin{tabular}{l rr rrr rrr}
        \toprule
        & \multicolumn{2}{c}{\textbf{\tracker[u]}} &  \multicolumn{3}{c}{\textbf{\statica[u]}} &  \multicolumn{3}{c}{\textbf{\dynamica[u]}} \\
        & \textit{Websites} & \textit{All Websites} & \textit{Websites} & \textit{All Websites}  & \textit{\tracker[u]} & \textit{Websites} & \textit{All Websites}  & \textit{\tracker[u]} \\
        \cmidrule(lr){2-3} \cmidrule(lr){4-6} \cmidrule(l){7-9}
        \rowcolor{lightgray} \textbf{Google} & 29,137 & 72.6\% & --- & --- & --- & 3,377 & 8.4\% & 11.6\% \\
        \textbf{Meta} & 11,309 & 28.2\% & 7,849 & 19.5\%   & 69.4\% & 7,049 & 17.6\% & 62.3\%  \\
        \rowcolor{lightgray} \textbf{Google $\cup$ Meta} & 29,363 & 73.1\% & --- & --- & --- & 8714 & 21.7\% & 29.7\% \\
        \bottomrule
    \end{tabular}
    \caption{Overview of all \tracker[s], and \tracker[s] with \statica and \dynamica for Google, Meta, and websites that have either (Google $\cup$ Meta). The denominator for ``All Websites'' percentages is 40,150.}
    \label{tab:dataset_overview}
\end{table*}

In this section, we briefly introduce our results and then proceed with an overview of our dataset validation process. A more in-depth discussion of our results is reserved for Section \ref{sec: analysis}.

Out of the 42,481 websites visited, we visited 40,150 (94.51\%) successfully. Unsuccessful visits can be attributed to several factors, including inaccessible domains, bot detection, and general unreliability. To reduce the likelihood of these errors, we implemented retry logic, re-visiting each website until a \dynamica event was detected or after the third visit. Google \dynamica and Meta \dynamica were considered independent events, i.e. even if a website had an instance of \dynamica from one of the two trackers on the first try, we still made additional attempts for the other tracker. All of our further analysis is based on only the 40,150 websites that we were able to successfully visit. Our analysis considers whether a website ever had a \tracker or \dynamica in \emph{any} of the scrapes over the three-month period of data collection. There is a possibility that a tracker configuration was altered between multiple visits, but we do not make such a distinction in our analysis. 

We observed the following marginal gains from each retry. For \gpixel \tracker[s], the first visit discovered 85.4\% of our total \tracker count, the second visit found an additional 13.67\%, and the third visit identified less than 1\%. We found a similar pattern for \mpixel \tracker[s]. The first visit discovered 91.28\%, the second visit 8.09\%, and the third visit less than 1\% of new observed \tracker[s]. For \gpixel \dynamica, we found that the first visit contributed 85.35\% of our total, the second visit 11.91\%, and the third visit 2.75\%. For \mpixel \dynamica, we found the first visit contributed 89.64\%, the second 7.48\%, and the third 2.88\%. Therefore, we believe further retries would not have led to a meaningful increase in either \tracker[] or \dynamica numbers.

Even with retries, by the nature of large-scale data scraping, many factors may have impacted the accuracy of our measurements. To quantify this impact, we performed a number of manual validations that address the following questions: (i) \textit{How many \tracker[s] and \dynamica[s] are we missing?} and (ii) \textit{How accurate is our \dynamica detection methodology as a proxy for measuring \statica?}

Additionally, as mentioned in Section \ref{sec:methodology}, we chose not to interact with cookie consent banners. In order to understand the impact this had, we performed an additional validation to answer the question: (iii) \textit{How many more \tracker[s] and \dynamica[s] would we detect if we accepted or rejected cookies?} 

We answered these validation questions by manually investigating samples of websites. Unlike our automated scraping, all validation was done on local machines by the paper's authors. The sample sizes used in our validation are based on population counts presented in Table~\ref{tab:dataset_overview}. Every sample referenced in this section was selected uniformly at random from the relevant population, and its size was calculated using a population proportion $\hat{p}= 0.5$ (the worst case) to achieve a 95\% confidence interval. 

\subsubsection{How many \tracker[s] and \dynamica[s] are we missing?}
\label{sec:tracker-validation}
To measure the extent to which we under-counted the number of \tracker[s,] we performed manual validation on two samples: websites with no detected \mpixel \tracker[s] and websites with no detected \gpixel \tracker[s.] Note that we considered \mpixel and \gpixel independent of each other, so each sample may have included instances of the other. We did not explicitly measure false positives.
Given that we identified \statica and \dynamica from parsed network events, we considered false positives to be unlikely.

For \gpixel[s], we visited a sample of 372 out of the 11,013 websites with no detected \gpixel \tracker[].
Of the 345 reachable websites, 91.6\% (316) were true negatives, and 8.4\% were false negatives. Similarly, out of the 28,841 websites without a detected \mpixel \tracker, we manually visited a sample of 380 websites. We successfully reached 368, of which 95.7\% (352) were true negatives and 4.3\% (16) were false negatives.

Although we do not know the exact reason for each missed \tracker[], we were able to identify some possible reasons through our manual validation. We observed instances where trackers loaded too slowly for our data collection infrastructure (i.e., our automation closed the website before the tracker had a chance to load) and instances where we were detected as a bot. It is also possible that some websites added \tracker[s] after we performed the data collection but before we performed manual validation.

We validated missed \dynamica events in a similar manner. For each website in the selected sample, we manually visited each website and injected and submitted a filled PII form. Although this task was similar to our form simulation task described in the methodology, it was performed manually on local machines, and thus, human judgment was used to determine how best to inject the form. 

To validate \dynamica events from \gpixel[,] we selected a sample of 379 websites from the 25,760 websites with \gpixel installation but no \dynamica[.] We successfully reached 364 websites but found that 13 of them did not have a \gpixel at the time of our validation, so we excluded them from our analysis. Of the remaining 351 websites, only 5 (1.4\%) were false negatives.

Similarly, for Meta, out of the 4,260 websites with \mpixel but no detected \dynamica[,] we investigated a sample of 353 websites. Of the 343 websites successfully visited, 19 websites did not have a \mpixel at the time of validation. Out of the remaining 324 websites, 31 (9.6\%) were false negatives.

We notice a higher percentage of missed \dynamica with \mpixel than \gpixel[.] This is possibly explained by the fact that far more \mpixel[s] in general are configured to collect data than \gpixel[s]. Although we cannot be precisely sure why we are under-counting, it could be attributed to the same reasons we observed missed \tracker[s]. In addition, we note that automated form injections can be challenging and, given the diversity of website designs, there is the potential that some of our injections failed, thus contributing to our under-counting.

We generally found that a subset of the false negatives could be attributed to cookie consent or other dialog boxes that blocked page load (17.3\%), bot detection (8.6\%), and websites we had not successfully visited (4.9\%).

\subsubsection{\dynamica[u] as a proxy for \statica[u]}

As described in Section \ref{sec: analysis-types}, we computed two different measures for \mpixel[s]: form data \textbf{configuration} and form data \textbf{collection}. We proceed by explaining the relationship between those two measures and show that collection is a suitable proxy for configuration. 

For each website with detected \statica, we looked at corresponding instances of detected \dynamica to validate that websites that we categorized as configured to collect data did have a measurable instance of data collection. Of the 7,849 websites with \statica, 89.8\% had \mpixel \dynamica, detected by inspecting network events; the remaining 10.2\% were false negatives. We further validated that specific PII fields (email, phone number, etc.) in the \statica were the same fields identified in the \dynamica network events.

To understand why 10.2\% of websites had \statica and no \dynamica, we took a uniformly random sample of 260 websites from the population of websites where our \statica and \dynamica results disagreed. After removing websites that either no longer had a \mpixel or no longer had \statica, we evaluated a sample of 219 websites. We found that 36.1\% of those websites did in fact have \mpixel \dynamica. We also observed that 17.8\% of our sample websites had multiple \mpixel[s] installed, where at least two trackers had \statica discrepancy (i.e. at least one \mpixel had \statica and at least one did not). As mentioned in Section~\ref{sec:qualitative}, this may lead to unpredictable \dynamica behavior, which could have contributed to our error rate. Assuming independent samples, we believe that for around a quarter of the websites that had \statica but no \dynamica, the error can be attributed to errors in our \dynamica methodology described above, rather than errors in our \statica analysis. 

In the opposite direction, we found that only 2 websites with detected \mpixel \dynamica did not have a detected \statica (0.28\%). 

We conclude that \dynamica detected based on injected forms is highly correlated with \statica and thus can be used as a suitable proxy to draw conclusions about website tracker configurations. In the following analysis, we will use \dynamica as a proxy for both \mpixel and \gpixel \statica, enabling us to compare between the two since we do not have \statica[s] for \gpixel.

\subsubsection{Effect of Accepting or Rejecting Cookies on \tracker[u] and \dynamica[u]}
Our decision not to interact with cookie consent banners means that we were unable to measure trackers that do not load until a visitor accepts cookies. In order to determine the impact this might have had on our measurements, we took a random sample of 371 websites from our dataset with no detected \mpixel or \gpixel installation. From this sample, we excluded any websites that did not load or had been erroneously classified and identified 146 websites with cookie consent banners. We accepted all cookies on these websites and ran the same process of data collection and analysis described above. 

We found that 43.8\% of websites had \mpixel \tracker[s] after accepting cookies, 24.0\% had \mpixel \statica[s] and 20.5\% had \mpixel \dynamica. These percentages are higher for \tracker[], and only slightly higher for \statica and \dynamica compared to our overall dataset. We found \gpixel \tracker[s] 71.9\% of the time and \gpixel \dynamica 8.2\% of the time, which closely matches what we saw in our overall dataset, which is more thoroughly discussed in Section~\ref{sec: analysis}.

We used the same sample to measure what would have happened had we rejected cookies. After excluding websites that either did not allow cookies to be rejected or required a subscription in the absence of cookies, we were left with 129 websites to test. We rejected all cookies and found that none of the websites had \mpixel \statica or \mpixel \dynamica. We found \gpixel trackers on 9.3\% of websites but no instances of \gpixel \dynamica.

Since this sample was uniformly chosen, we believe with 95\% confidence (assuming independence) that these percentages would apply to all websites with neither tracker (10,787) had we accepted or rejected cookie consent banners.

\subsubsection{Validation Takeaway}
Based on our validation, we conclude that our measurements likely provide a lower bound on the number of \tracker[s], \statica[s], and \dynamica[s] that exist on websites we visited. While all validations had an error rate no higher than 10\% (except websites that had \statica and no \dynamica), the bound is tightest for \gpixel \tracker[s], which had the lowest rate of miss-classification. We believe over-counting is unlikely, as we only classify a website as having \dynamica if PII is identified in one of the dynamic URLs in Table \ref{tab:collection-urls}, which specifically route to Meta and Google domains. However, it is possible websites have modified tracker configurations to no longer collect data since our initial visit.

%% file: table_subsets.tex
\begin{table}[ht]
    \centering
    \small
    \begin{tabular}{lrrr}
        \toprule
        \multicolumn{1}{c}{\multirow{2}{*}{\textbf{Subset}}} & \multicolumn{1}{c}{\multirow{2}{*}{\textbf{Websites}}}  & \multicolumn{2}{c}{\textbf{\dynamica[u]}} \\
        & & \textbf{Google} & \textbf{Meta} \\
        \midrule
        \rowcolor{lightgray} \textbf{\gpixel $\cap$ \mpixel} & 11,083 & 21.7\% & 62.7\% \\
        \textbf{\gpixel $\cap$ $\neg$\mpixel} & 18,054 & 5.4\% &  \textemdash \\
        \rowcolor{lightgray} \textbf{$\neg$\gpixel $\cap$ \mpixel} & 226  & \textemdash  & 44.2\%\\    
        \bottomrule
    \end{tabular}
    \caption{Breakdown of \dynamica for different websites with both trackers, websites with only \gpixel, and websites with only \mpixel.}
    \label{tab:breakdown_subsets}
\end{table}

%% file: ethics.tex
\section{Ethical Considerations}

\noindent{\textbf{Data Collection.}}
The PII used to submit forms on each website was placeholder data generated by the authors of this paper. No real PII was collected. Further, since our form was not tied to any website infrastructure, our placeholder data is much less likely to pollute any website’s real advertising ecosystem or even be sent to the website directly.

\noindent{\textbf{Disclosures.}}
We disclosed observed form data collection to Meta, Google, and 121 websites we determined to be in health and finance verticals. We clarified in our disclosure letter (Appendix Section~\ref{sec:notification-template}) that we did not collect or observe any real user data but simply the potential for data collection.

%% file: discussion.tex
\section{Discussion}
\label{sec:discussion}

PII form data collection primarily exists to enable the identification of a specific individual for cross-device or, more broadly, cross-context tracking. Since people often own multiple devices, cross-context tracking is essential for creating a more complete profile of a person's activity, which often increases the value of a website visitor in targeted advertising auctions across third-party marketing tools. Our methodology, which analyzed PII form data collection from multiple perspectives, offers a deeper understanding of the mechanics behind this PII data collection than prior work.

We found that website administrators face several challenges when configuring trackers. First, the tracking documentation provided by third parties to website administrators serves a dual role as a product guide and marketing material. As a result, it contains instances of marketing language that focus on the benefits of the data collection feature at the expense of adequately explaining the details of how it works. For example, Meta introduces automatic data collection as something that can ``help you optimize your Meta ads to drive better results''~\cite{facebookHash}. Similarly, Google states that ``data collected helps customers understand their users' needs''~\cite{googleUserNeeds}. Even more concerning, we uncover that both Meta and Google assert hashing PII as a legitimate privacy technique to website administrators. Although, from an economic lens it is understandable why Google and Meta continue to treat hashing as an adequate privacy solution, both the US FTC and the research community have repeatedly debunked hashing as ineffective for privacy.

Our website measurements indicate that there is likely a divergence in the frequency at which website administrators configure trackers to collect PII form data. Website administrators appear more likely to configure \mpixel to collect PII form data than \gpixel. This aligns with our analysis that Meta's configuration UI flow forces the website administrator to enable or disable automated PII form data collection and recommends enabling this feature, whereas it is possible to complete the \gpixel configuration flow without encountering a PII form data collection prompt.

As we demonstrated, and to Google and Meta's credit, they block automated PII form data collection on websites self-declared as belonging to the health or finance vertical. However, based on our website measurements, we find that websites clearly in these categories, such as drug and alcohol rehab centers and banks, have in fact installed \mpixel[s] and \gpixel[s] to collect PII form data. This indicates that some websites in regulated industries may be incorrectly declaring their verticals. We contacted 121 websites that fell into this category to notify them of a PII form data collection configuration (although we are not explicitly verifying actual collection of customer data). Four websites have followed up with an intention to review website configurations. After our disclosure, NAMI made the necessary modifications and is no longer configured
for form data collection.

There are several tactics that may reduce form data collection, particularly on sensitive websites. Website administrators may need to remain vigilant about how trackers are deployed, especially when relying on an external contractor or other third party to manage tracker configurations. Additionally, there are technical defenses available to website visitors, like ad blockers, that can help prevent form data collection. However, they may degrade website functionality by triggering ad-blocker detection that withholds website content until trackers are enabled or break parts of the website by accidentally blocking key scripts. As an alternative, it is possible to proactively warn website visitors that a tracker present on the page has been configured to collect form data. We created a proof-of-concept Chrome extension, included in the paper's artifacts, that can analyze \mpixel tracking code loaded by a website and notify website visitors which PII fields, if any, will likely be sent to Meta before they fill out any form. We open-source this extension with the hopes that the community can expand it to include notifications for other popular trackers, such as \gpixel. Unfortunately, this mitigation is imperfect since it both places a burden on the website visitor, who may be unable to access services without interacting with a web form, and does not detect any server-side PII data collection mechanisms.

Ultimately, decreasing the use of PII form data collection requires a diverse and comprehensive set of interventions from industry, government, and the research community. While data regulation does exist in the United States, this research demonstrates that it is insufficient without further enforcement actions, particularly in improving the documentation and configuration interfaces provided by Meta and Google. Further, additional regulation in the United States is likely necessary to enforce alternatives to hashing that provide strong privacy, which would likely reduce ad revenue by making it challenging to re-identify a person. Although this study exclusively visited websites from the United States, it has applications to other jurisdictions. For example, health data is also a protected category under the European Union's GDPR~\cite{gdpr}. Beyond regulation, we need solutions that enable privacy-preserving methods of profile linking and targeting to lessen the economic impact on advertisers and advertising networks while weening them off their fire hose of PII.

%% file: appendix.tex
\section{UI and Network Screenshots}
\label{sec:appendixa}
Here we include figures that demonstrate the configuration interfaces and network events discussed in Section~\ref{sec:quantitative}.

\begin{figure}[ht]
        \centering
        \includegraphics[width=\columnwidth]{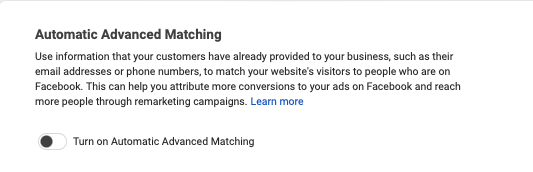}
        \caption{Meta UI - Data Collection Prompt}
        \label{fig:meta-UI}
\end{figure}

\begin{figure}[ht]
        \centering
        \includegraphics[width=\columnwidth]{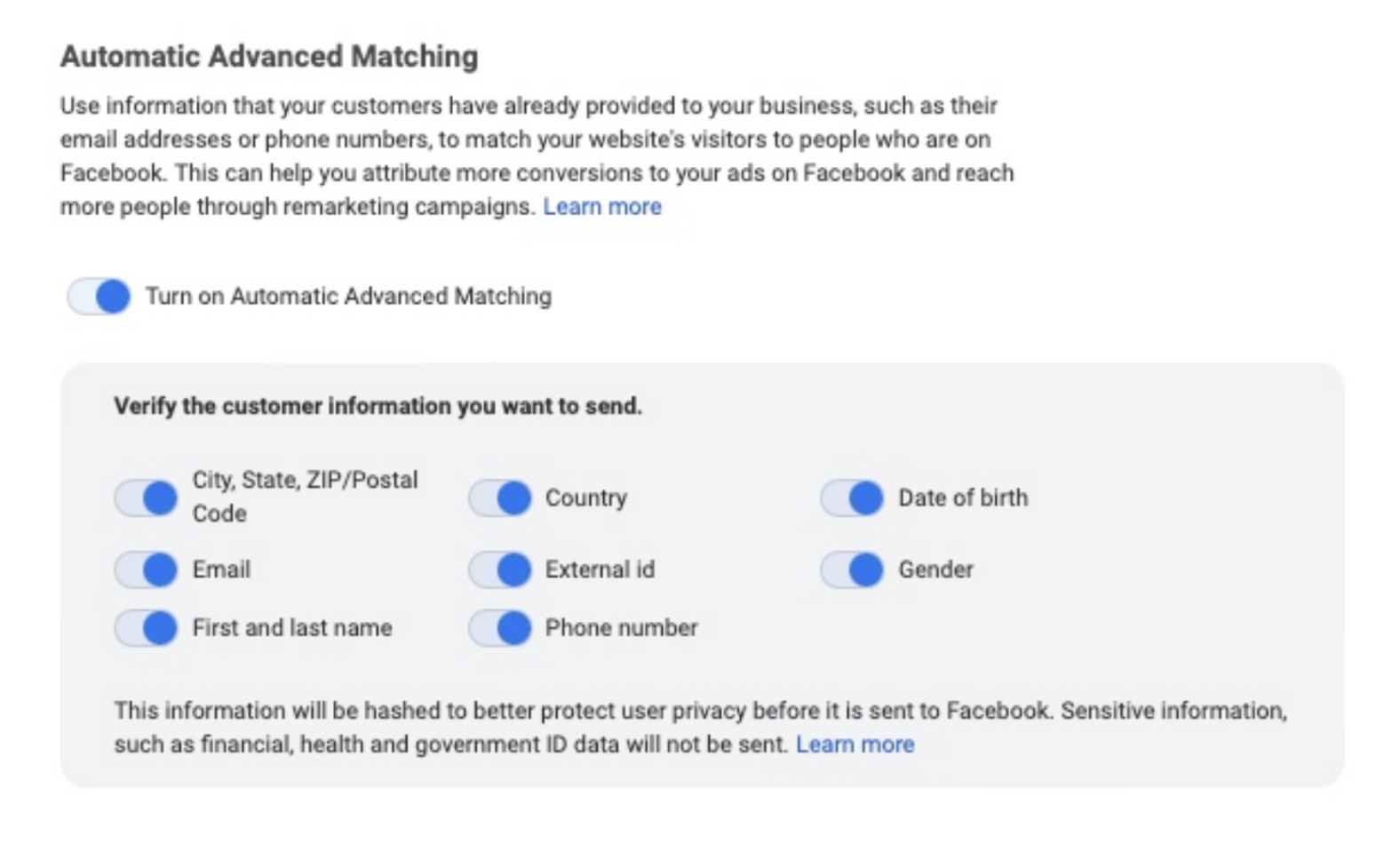}
        \caption{Meta UI - Data Collection Fields}
        \label{fig:meta-auto-select}
\end{figure}

\begin{figure}[ht]
        \centering
        \includegraphics[width=\columnwidth]{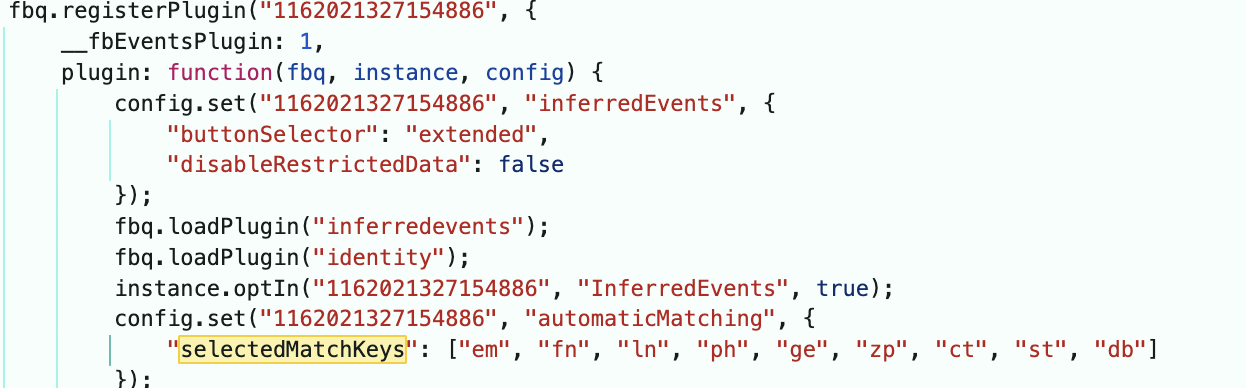}
        \caption{Meta Data Collection in the Code Configuration}
        \label{fig:meta-config}
\end{figure}

Figures \ref{fig:meta-UI} and \ref{fig:meta-auto-select} illustrate how a website administrator is presented with form data collection in the setup flow of \mpixel[,] first with a UI prompt to toggle on, and then with the resulting auto-selection of all available PII fields after taking the prompt.
Figure~\ref{fig:meta-config} provides a snapshot of where these options are reflected in the generated configuration code of \mpixel[.]

\begin{figure}[ht]
        \centering
        \includegraphics[width=\columnwidth]{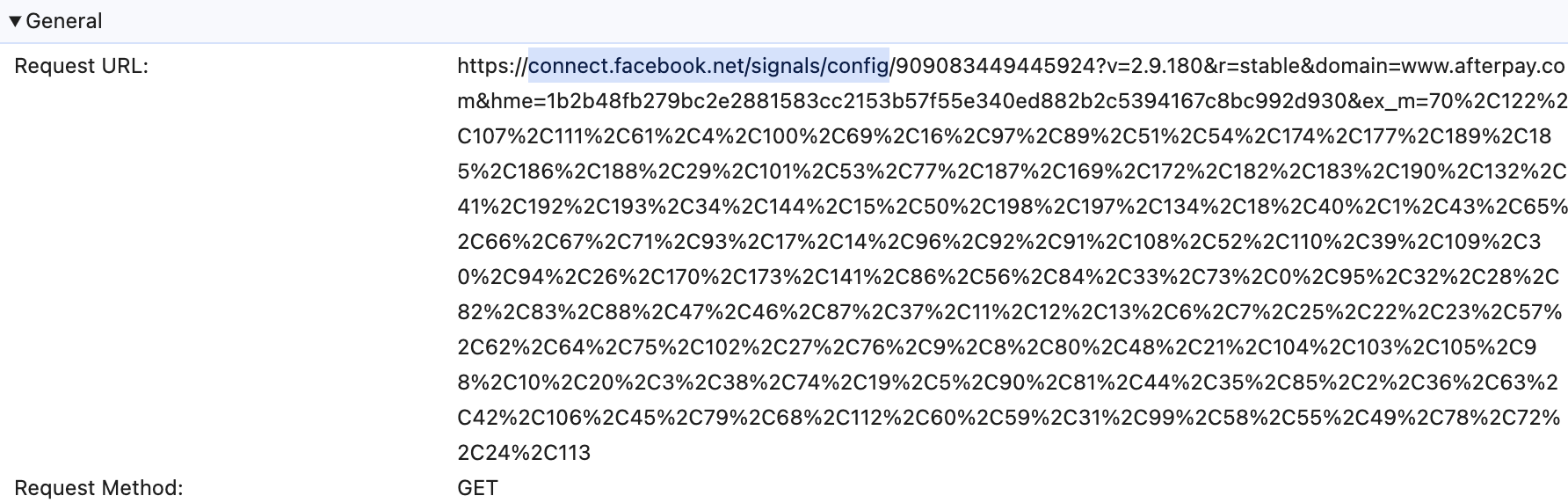}
        \caption{GET request for Meta Pixel}
        \label{fig:meta-get-request}
\end{figure}

\begin{figure}[ht]
        \centering
        \includegraphics[width=\columnwidth]{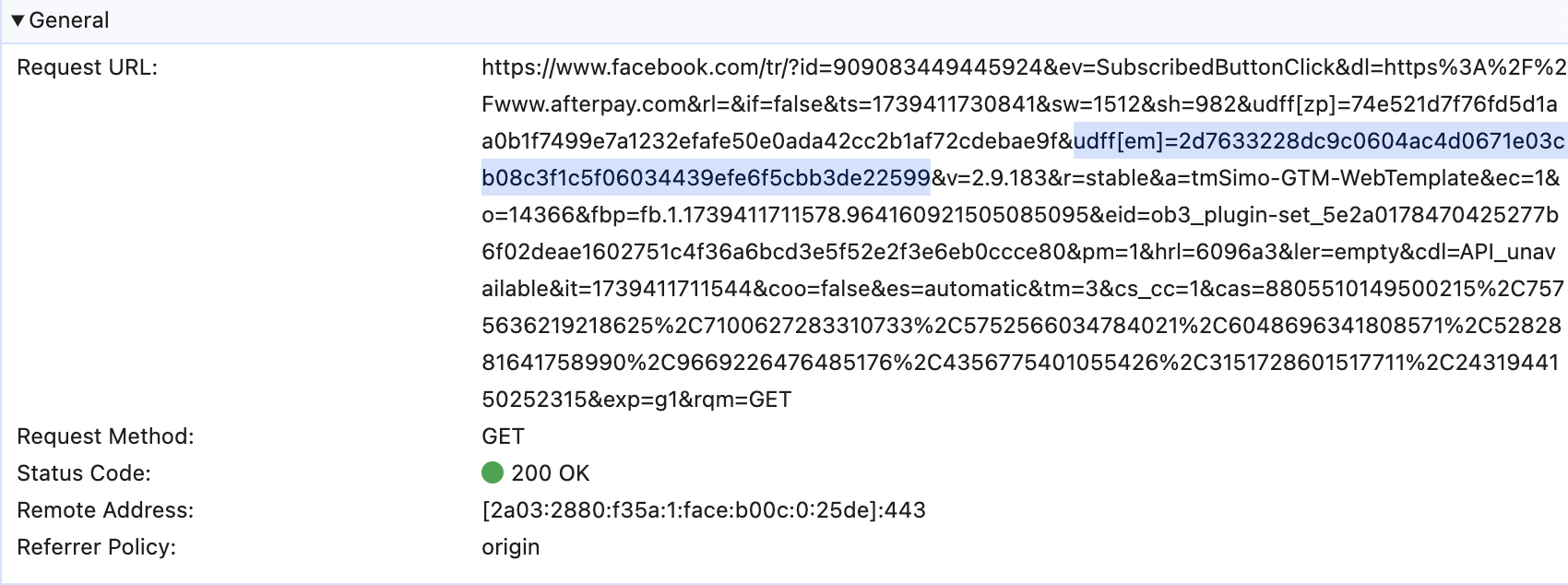}
        \caption{Form Data Collection Event for Meta Pixel}
        \label{fig:meta-fdc-event}
\end{figure}

\begin{figure}[ht]
        \centering
        \includegraphics[width=\columnwidth]{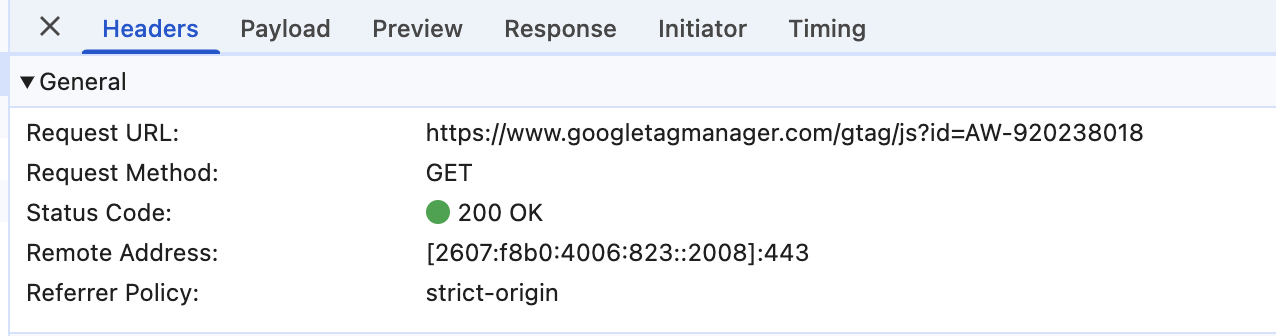}
        \caption{GET request for Google Tag}
        \label{fig:google-get-request}
\end{figure}

\begin{figure}[h]
        \centering
        \includegraphics[width=\columnwidth]{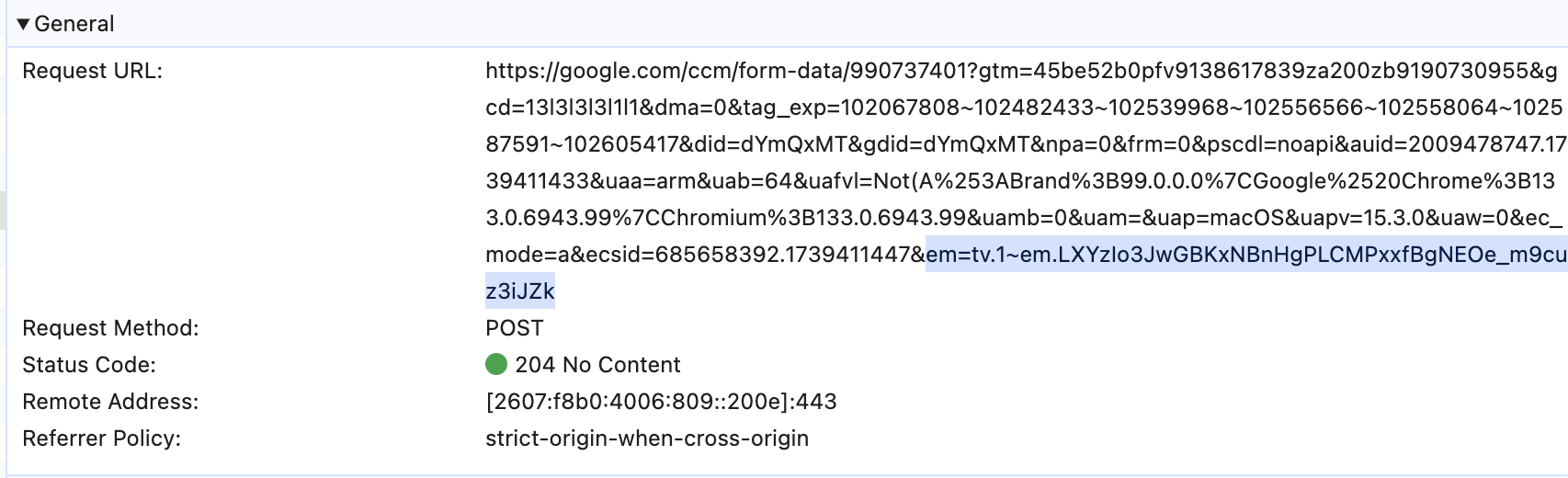}
        \caption{Form Data Collection Event for Google Tag}
        \label{fig:google-fdc-event}
\end{figure}

Figures~\ref{fig:meta-get-request}, \ref{fig:meta-fdc-event}, \ref{fig:google-get-request}, and \ref{fig:google-fdc-event} illustrate the URLs and hashed PII identified by our parser to label website installations and form data collection.

\section{Regression Analysis Tasks}
\label{sec:appendix-regression}
\input{logistic_regression_analysis}
To assist us with providing these insights, we performed two separate Logistic Regression analysis tasks. Table~\ref{tab:logistic_regression} provides an overview of both of them; in the first, we trained a model to predict whether a website performs \dynamica through Meta. We trained this dataset on only websites that already have a \mpixel. We used as features four boolean variables; the first is true when a website has a \gpixel, the second when a website has Google \dynamica, the third when a website is in the Health vertical, and the fourth when it is in the Finance vertical.

The second model was trained on only websites that have a \gpixel, and trained on three features; the first feature is true when the website has a \mpixel, and once more the two features that signify whether the website belongs in the sensitive Health or Finance vertical, respectively. We do not use for this model a feature to signify Meta \dynamica because there was a very high correlation between this feature and the existence of a \mpixel (0.71 Pearson's correlation coefficient). We kept the feature that was a better predictor.

It is important to acknowledge that our models are a relatively weak fit; the former has a pseudo R-squared of 0.0721 and the latter has a pseudo R-squared of 0.0825. This means that their explanatory power is limited. However, we can still draw some useful insights from them especially if we can combine them with our measurements.

\subsection{Meta Form Data Collection Model Results}
\label{sec:appendix-regression-meta}

The upper half of Table~\ref{tab:logistic_regression} presents the results for our Meta \dynamica model. When a website has a \gpixel then it is 1.903 times more likely, and when it has Google \dynamica it is 1.699 times more likely to have Meta \dynamica[.] We also see that the odds ratios for Health (0.206) and for Finance (0.118) suggest that it is 79.4\% less likely for a website to have Meta \dynamica when it belongs in the Health vertical, and 88.2\% less likely when it belongs to the Finance vertical.

\subsection{Google Form Data Collection Model Results}
\label{sec:appendix-regression-google}

Regarding Google \dynamica[,] we see in the lower half of Table~\ref{tab:logistic_regression} that according to our model, when a website in our training dataset has a \mpixel, it is 4.839 times more likely to have Google \dynamica[.] 
In addition, we note that belonging to either Health or Finance is not a statistically significant feature for  predicting Google \dynamica[.]

\section{Notification Template}
\label{sec:notification-template}
We identified 121 websites that had form data collection configured and we conservatively believed to be in a Health or Finance vertical. We contacted them through emails found, in order of preference, as the technical email associated with the domain, an IT support contact, a general contact, or a PR contact email. We used the following template for notification.

To Whom It May Concern,

I am [name and affiliation]. Our research team is studying third-party tracker configurations on [health | finance] websites. Based on the products and services detailed on your website, we believe you can be classified as a [health | finance] website.

We believe that your website includes a [Google | Meta] tracker (with ID [TRACKER ID]) that is configured to collect [emails | visitor data, including emails] and send them to [Google | Meta]. This configuration may be set up through the [Google | Meta] user interface. However, websites categorized as [health | finance] are not permitted to use this collection feature by [Google | Meta]. Therefore, we believe trackers included in your website have been mis-categorized during setup.

As it is currently configured, the tracker(s) may collect data and share it with [Meta | Google] when a user visits your website and fills out a form, such as a login, contact, or subscription form. We did not specifically verify that your website has a form requesting this information from visitors, and all of our testing was done with fabricated data; we did not view any real customer information. Testing was completed between September and November 2024.

Sincerely,
[name]
[contact email]

%% file: logistic_regression_analysis.tex
 \begin{table}[ht]
        \centering
        \small
        \begin{tabular}{p{1pt}lrrr}
            \toprule
            &\textbf{Feature} &  \multicolumn{1}{c}{\textbf{OR}} &  \multicolumn{1}{c}{\textbf{$p$-value}} &  \multicolumn{1}{c}{\textbf{CI}} \\
            \midrule
            \multirow{4}{*} {\rotatebox{90}{\textbf{Meta}}} & Has \gpixel &  1.903 &  0.000 & [1.443, 2.512] \\
            & Google \dynamica[u] &  1.699 &  0.000 & [1.533, 1.885] \\
            & Is Health &  0.206 &  0.000 & [0.180, 0.237] \\
            & Is Finance &  0.118 &  0.000 & [0.094, 0.147] \\
            \midrule
            \multirow{3}{*} {\rotatebox{90}{\textbf{Google}}} & Has \mpixel &  4.839 &  0.000 & [4.473, 5.233] \\
            & Is Health &  0.952 &  0.457 & [0.835, 1.084] \\
            & Is Finance &  1.086 &  0.376 & [0.905, 1.305] \\
            \bottomrule
        \end{tabular}
        \caption{Odds Ratios (OR), $p$-values, and Confidence Intervals (CI) from our Logistic Regression Analyses: \\ (i) With \mpixel \dynamica as dependent variable; trained on all websites that have \mpixel (Pseudo R-squared: 0.0721). Results suggest that a website is more likely to have  \mpixel \dynamica when it has \gpixel and \gpixel \dynamica, and less likely if it belongs to Health or Finance verticals.\\ (ii) With \gpixel \dynamica as dependent variable; trained on all websites that have \gpixel (Pseudo R-squared: 0.0825). Results suggest that a website is more likely to have \gpixel \dynamica when it has a \mpixel. \\ All features are boolean variables -- True if a website has the property. }
        \label{tab:logistic_regression}
    \end{table}